\documentstyle[]{l-school}
\begin{document}
\markboth{St. D. G{\l}azek}{RENORMALIZATION OF HAMILTONIANS}
\title{Renormalization of Hamiltonians}
\author{Stanis{\l}aw D. G{\l}azek}
\institute{Institute of Theoretical Physics, Warsaw University\\
ul. Ho\.za 69, 00-681 Warsaw, Poland}
\maketitle
\begin{abstract}
A matrix model of an asymptotically free theory with a bound state is
solved using a perturbative similarity renormalization group for
hamiltonians.  An effective hamiltonian with a small width, calculated
including the first three terms in the perturbative expansion, is projected
on a small set of effective basis states.  The resulting small
hamiltonian matrix is diagonalized and the exact bound state energy is
obtained with accuracy of order 10\%.  Then, a brief description and an
elementary illustration are given for a related light-front Fock space
operator method which aims at carrying out analogous steps for
hamiltonians of QCD and other theories.
\end{abstract}
\section{INTRODUCTION}
This lecture has two aims.  The first aim is to show a simple example of
a new kind of calculation of effective hamiltonians, based on the
perturbative similarity renormalization group.  \cite{GW1} \cite{GW2}
The second aim is to show how one can generalize the simple example and
start systematic perturbative calculations for quantum field theoretic
hamiltonians in the light-front Fock space.  

Although the methods we present are quite general, the main motivation 
came from QCD. QCD is asymptotically free and its perturbative running coupling
constant grows at small momentum transfers beyond limits.  This rise
invalidates usual perturbative expansions in the region of scales where
the bound states are formed.

Ref.\cite{W3} outlined a light-front hamiltonian approach to this problem
in QCD, using the perturbative similarity renormalization group.
Independently, Wegner \cite{WEG} proposed a flow equation for
hamiltonians in solid state physics. He introduced an explicit 
expression for the generator of the similarity transformation which 
leads to a Gaussian similarity factor of a uniform width.

Wilson and I have solved numerically a simple matrix model to gain 
quantitative experience with the similarity scheme using Wegner's 
equation. \cite{WG} We also made perturbative studies. \cite{GW3} 
This lecture is based on those works in the part describing the model.
The remaining part contains an outline of how one can attempt to make 
similar steps for light-front hamiltonians in quantum field theory 
using creation and annihilation operators.\cite{G} 
\section{MODEL}
Consider a quantum theory which is characterized
by a large range of energy scales as measured by certain $H_0$.  QCD has
this feature. It extends in energies from $\infty$ (asymptotic
freedom) down to the infrared energy region. We represent the 
theory by a model with a hamiltonian $H = H_0 + H_I$ acting in a 
space spanned by a finite discrete set of nondegenerate 
eigenstates of the hamiltonian $H_0$,
$$ H_0 |i> \quad = \quad E_i |i> \, . \eqno(2.1) $$
\noindent Matrix elements of the interaction are assumed to be
$$ <i| H_I |j> \quad = \quad - g \sqrt{ E_i E_j } \, . \eqno(2.2) $$
\noindent $g$ is a dimensionless coupling constant.  

We choose $E_i = 2^i$ and $M \leq i \leq N$.  $M$ is large and negative 
and $N$ is large and positive. We use $M=-21$ and $N=16$ in our numerical 
example.  Let the energy equal 1 correspond to 1 GeV.  Then, the ultraviolet 
cutoff corresponds to 65 TeV and the infrared cutoff corresponds to 0.5 eV.

The same model can be alternatively derived by discretization of the
2-dimensional Schr\"odinger equation with a potential of the form a
coupling constant times a $\delta$-function. \cite{J}

For $g > 1/38$, the hamiltonian matrix has one
negative eigenvalue and 37 positive eigenvalues.  
$g$ is adjusted to obtain the negative eigenvalue equal
$-1$ GeV; $ g \sim 0.06$. This eigenvalue corresponds to
the s-wave bound state energy in the 2-dimensional 
Schr\"odinger equation.

We calculate effective hamiltonians, ${\cal H} \equiv {\cal
H}(\lambda)$, using the similarity renormalization group equations in
the differential form.  The effective hamiltonians are parametrized by
their energy width $\lambda$. The notion of the hamiltonian
width will become clear shortly. We use Wegner's flow equation
\cite{WEG}
$$ {d {\cal H} \over d\lambda^2} \quad = \quad
 - \, {1 \over \lambda^4 }
\left[\, [{\cal D}, {\cal H}],\, {\cal H}\, \right] \, ,
 \eqno(2.5) $$
\noindent with the initial condition
${\cal H}(\infty) = H$. The matrix ${\cal D}$ is  
the diagonal part of $\cal H$ with elements
$ {\cal D}_{mn} = {\cal H}_{mm} \delta_{mn}$.
Thus, ${\cal H}(\lambda)$ is a unitary
transform of $H$ and both have the
same spectrum (see Wegner's lecture in this volume).

Equation (2.5) can be approximately solved for a small
$g$ keeping only terms order $1$ and $g$. One obtains
$$ {\cal H}_{mn} \quad = \quad E_m \, \delta_{mn} \quad - \quad g
\sqrt{E_m E_n} \,\, \exp\left[ - (E_m - E_n)^2/\lambda^2 \right] \,\, .
\eqno(2.6) $$
\noindent Here, ${\cal D}_{mm} = (1-g) E_m$. The Gaussian factor 
of width $\lambda$ is the similarity function.  
This explains the notion of the hamiltonian width.  
Ref.  \cite{WG} demonstrated that the Wegner flow equation has a 
renormalization group interpretation.
\begin{figure}[b]
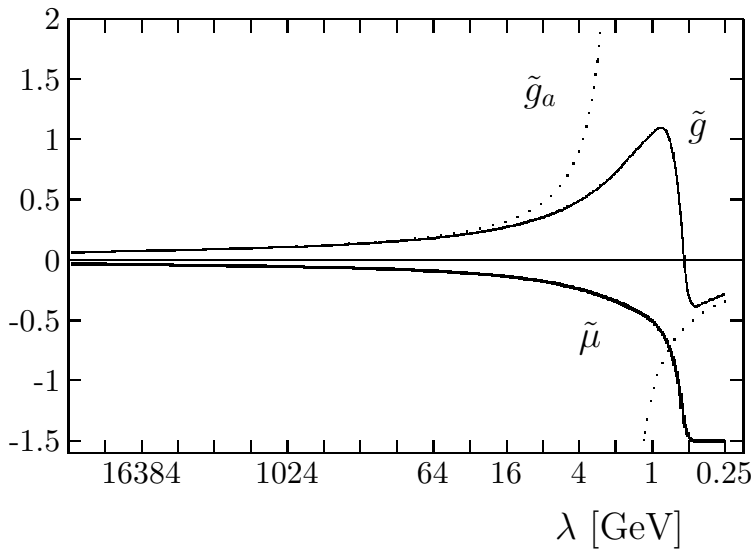

\begin{center}
\setlength{\unitlength}{0.240900pt}
\ifx\plotpoint\undefined\newsavebox{\plotpoint}\fi
\sbox{\plotpoint}{\rule[-0.200pt]{0.400pt}{0.400pt}}%

\end{center}
\vskip1cm
\caption{The approximate running coupling $\tilde g_a(\lambda)$ from Eq. (2.8)
and the exact running coupling $\tilde g(\lambda)$, plotted as functions of 
the effective hamiltonian width $\lambda$. The matrix element 
$\tilde \mu(\lambda) = {\cal H}_{-1,-1}(\lambda) - 0.5$ GeV is also plotted
to show the width range where the bound state eigenvalue appears on the
diagonal.}
\end{figure}
Including terms order $g^2$, we let $g$ depend on
$\lambda$ and we introduce $\tilde g (\lambda) \equiv \tilde g$. 
It follows from equations satisfied by the matrix elements 
${\cal H}_{mn}$ with 
the indices $m$ and $n$ close to $M$ that, neglecting small energies, 
$$ d \tilde g /d\lambda \quad = \quad - \, 
\tilde g^2 {d \over d\lambda} \sum_{\ell}
\exp{[-2 {E}_{\ell}^2 / \lambda^2]} \,\, , \eqno (2.7) $$
\noindent and $\tilde g(\infty) = g $. Analytic integration of Eq. (2.7) 
in the model gives, approximately,
$$ \tilde g_a (\lambda) \quad = \quad 
(1.45 \log{\lambda} - 0.9)^{-1} \,\, . \eqno(2.8)$$
\noindent $\tilde g_a(\lambda)$ grows when $\lambda$ gets smaller
and it exhibits the asymptotic freedom behavior: it is smaller for
more violent interactions (i.e. of wider range in energy). 
$\tilde g_a(\lambda)$ blows up to infinity for $\lambda \sim 1.9$ GeV. 
In this approximation, matrix elements of ${\cal H}$ for $E_m \sim E_n \ll
\lambda$ can be written as
$$ {\cal H}_{mn}(\lambda) 
\, = \, E_m \delta_{mn} \, - \, \tilde g_a(\lambda) \,
\sqrt{E_m E_n} \,\, \exp\left[ - [E_m - E_n]^2 / \lambda^2 \right]
+ corrections \,\, . \eqno(2.9) $$
\noindent Now, ${\cal D}_{mm}(\lambda) = [1 - \tilde g_a(\lambda)] E_m$. The 
energy order of low energy states is reversed when $\tilde g_a(\lambda)$
grows above 1.

The exact running coupling, $\tilde g(\lambda)$, is defined by writing
${\cal H}_{M, M+1}(\lambda) = - \tilde g(\lambda) \\ \sqrt{E_M E_{M+1}}$. 
Eq. (2.9) shows that $\tilde g_a = \tilde g$ for large $\lambda$. To find
$\tilde g$ for all values of $\lambda$, we solved Eq. (2.5) numerically.
\begin{figure}[b]
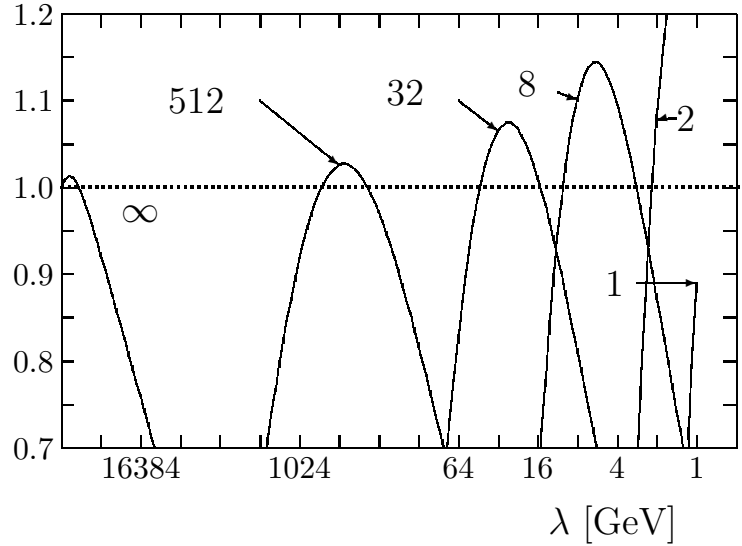

\begin{center}
\setlength{\unitlength}{0.240900pt}
\ifx\plotpoint\undefined\newsavebox{\plotpoint}\fi
\sbox{\plotpoint}{\rule[-0.200pt]{0.400pt}{0.400pt}}%

\end{center}
\vskip1.3cm
\caption{ The accuracy of the bound state eigenvalues obtained
from effective hamiltonians whose renormalization group flow with
the width $\lambda$ is
calculated expanding in powers of the effective coupling constant
$\tilde g (\lambda_0)$ and including terms order $1$,
$\tilde g (\lambda_0)$ and $\tilde g^2(\lambda_0)$.
The accuracy is given as ratio of the
bound state eigenvalue obtained by diagonalization of the effective
hamiltonian of width $\lambda$ to the exact value, $- 1$ GeV.
The curves correspond to the indicated values of
$\lambda_0$ (in units of GeV). The result of expansion in
the initial coupling $g$ is denoted by $\infty$.
The arrows show points where $\lambda = \lambda_0$.}
\end{figure}
Fig. 1. shows that the approximate solution blows up in the flow 
before the effective hamiltonian width is reduced to the scale 
where the bound state is formed.
That scale, order 1 GeV, equals $\lambda$ at which the bound state 
eigenvalue appears on the diagonal. The diagonal matrix element
is also shown in Fig. 1.

The key feature, visible in Fig. 1, is that the exact effective coupling 
constant does not grow unlimitedly. The similarity renormalization group
for hamiltonians provides a new option for investigating bound state
dynamics in asymptotically free theories. The question is how far down 
in $\lambda$ we can reach using perturbation theory instead of the exact 
solution. The answer is: down to 1 GeV in second order with 10\% accuracy.
This is illustrated in Fig. 2.

\begin{table}[t]
\begin{center}
\begin{tabular}{|c|c|c|c|}\hline
 window/whole   & $\tilde n = 2$ & $\tilde n = 1$ & $\tilde n = 0$ \\ \hline
$\tilde m = -8$ &         0.993  &         0.993  &          0.961 \\ \hline
$\tilde m = -5$ &         0.940  &         0.940  &          0.908 \\ \hline
\end{tabular}
\end{center}
\caption{Ratio of the bound state eigenvalue of the small window hamiltonian
with indices limited by $\tilde m$ and $\tilde n$, to the eigenvalue
of the whole effective hamiltonian at $\lambda = 1$ GeV calculated using
expansion up to second power in the running coupling $\tilde g(1 {\rm GeV})$.
0.993 corresponds to the absolute accuracy of the bound state eigenvalue
equal 12\% and 0.908 to 19\% (see the text).}
\end{table}
The remaining question of how small the space of states can be
on which one can project the narrow effective hamiltonian and reproduce
the bound state eigenvalue by diagonalization of the projected matrix,
is answered in {\it Table I}. The eigenvalue of the whole
${\cal H}(\lambda = 1 {\rm GeV})$ is equal $ - 0.8902$ GeV 
instead of $ - 1$ GeV. Window matrices with energy range order 1 GeV 
reproduce the same result with accuracy given in {\it Table I}. This is 
encouraging to pursue a similar approach to QCD. 
\section{FOCK SPACE METHOD}
The model study shows that the perturbative similarity renormalization group
allows a calculation of a small width effective hamiltonian, which can be
projected on a small space of states. The small hamiltonian can be solved
exactly and the bound state eigenvalue of the full theory is obtained with
10\% accuracy. The question is how to repeat these steps in quantum field
theory.

The method we propose \cite{G} is based on the idea that one can unitarily
transform the creation and annihilation operators, i.e.
$$   a^\dagger_{\lambda} = U_\lambda a^\dagger_\infty
U^\dagger_\lambda \, ,\eqno (3.1) $$
\noindent and the same for $a$'s. 
$a^\dagger_\infty$ and $a_\infty$ appear in the
initial hamiltonian $H$. We call them ``bare''. $a^\dagger_\lambda$ 
and $a_\lambda$ appear in the effective hamiltonian ${\cal H}_\lambda$.
They create and annihilate effective particles. In a way, $U_\lambda$ 
is analogous
to the Melosh transformation in the case of quarks. However, we are building
the transformation using the similarity renormalization group idea, the
transformation is fully dynamical and it can be applied to other particles
than quarks, too.

The effective hamiltonians satisfy the equation
$$ {d \over d\lambda} {\cal H}_\lambda = [ {\cal H}_\lambda, {\cal
T}_\lambda] \, ,\eqno (3.2) $$
\noindent where $ {\cal T}_\lambda = U^\dagger_\lambda dU_\lambda/d\lambda$.
The unitary transformation generator ${\cal T}$ is constructed so
 that the effective hamiltonians have width $\lambda$
in the relative momentum transfer,
$$ {\cal H}_\lambda = F_\lambda[{\cal G}_\lambda].  \eqno (3.3) $$
The operation $F_\lambda$ on the interaction terms 
${\cal G}_\lambda$, inserts the similarity factors, $f_\lambda$.
They are most easy to think about as form factors in the interaction vertices.
The smaller is $\lambda$ the softer are the interactions and the effective
particles get more dressed.

Following the general idea of the similarity scheme \cite{GW2},
one can find the equation satisfied by the vertex operators in
the effective hamiltonians \cite{G}, i.e.
$$     {d \over d\lambda} {\cal G}_\lambda =
\left[ f_\lambda {\cal G}_{2\lambda}, \left\{ {d \over d\lambda}
(1-f_\lambda) {\cal G}_{2\lambda} \right\}_{{\cal G}_{1\lambda}} \right] .
\eqno (3.4) $$
\noindent ${\cal G}_\lambda = {\cal G}_{1\lambda} + {\cal G}_{2\lambda}$,
${\cal G}_{1\lambda}$ is the $a^\dagger a$ part of the hamiltonian and
${\cal G}_{2\lambda}$ is the remaining part which changes
momenta of the individual particles. The curly bracket with subscript
${\cal G}_{1\lambda}$ denotes the similarity energy denominator
factor.

An elementary example illustrates how it works in Yukawa theory
which is defined by the following initial hamiltonian
$$ H_Y = \int dx^- d^2 x^\perp \left[ \bar \psi_m \gamma^+
{-\partial^{\perp 2} +
m^2 \over 2 i\partial^+} \psi_m +
{1\over 2}\phi (-\partial^{\perp 2} + \mu^2 )\phi \right.$$
$$ \left. + g \bar \psi_m \psi_m \phi + g^2\bar \psi_m \phi {\gamma^+
\over 2i\partial^+}\phi \psi_m \right]_{x^+=0}. \eqno(3.5) $$

\noindent The one particle energy is obtained in the form,
$$  {\cal G}_{1\,meson\,\lambda} =  \int[k] { k^{\perp 2} +
\mu^2_\lambda \over k^+ } a^\dagger_k a_k . \eqno (3.6) $$
\noindent In second order perturbation theory in the coupling constant $g$,
Eq. (3.4) implies
$$ {d \mu^2_\lambda \over d\lambda} = g^2 \int[x\kappa]
{d f^2(z^2_\lambda) \over d\lambda}
{ 8(x-{1\over 2})^2{\cal M}^2 \over {\cal M}^2 -\mu^2 }
r_\epsilon(x,\kappa) . \eqno (3.7) $$
\noindent Here ${\cal M}^2 = (\kappa^2 + m^2) / x(1-x) $ and
$r_\epsilon(x,\kappa)$ denotes the regularization factor which is an analog
of the number $N$ in the matrix model. The similarity function 
$f^2(z_\lambda^2)$ can be made 
as simple as, for example, 
$\theta(\lambda^2 + 3\mu^2 - {\cal M}^2)$. In this case,
integration of Eq. (3.7) gives the following effective meson mass term
$$ \mu^2_\lambda = \mu_1^2 + {\alpha \over 48\pi} \left[\lambda^2
-\lambda_1^2 + (\mu^2 -6m^2) \log{\lambda^2 \over \lambda_1^2}\right] +
\mu^2_{conv}(\lambda,\lambda_1) + o(g^4) . \eqno (3.8) $$
\noindent $\mu^2_{conv}(\lambda,\lambda_1)$ denotes a finite term which 
has a limit when $\lambda \rightarrow \infty$. It equals 0 for $\lambda = 
\lambda_1$. $\mu_1$ is the effective meson mass in the hamiltonian
${\cal H}(\lambda_1)$. In the second order calculation, it is equal to 
the physical meson mass if $\lambda^2_1 \leq 4m^2 - 3\mu^2$.

The reason I show this example is that one can do similar calculations
for other terms in the effective hamiltonians. \cite{G}
For example, in second order perturbation theory, effective
interactions between quarks are partly similar to the results obtained by
Perry and his collaborators. \cite{Perry} \cite{BP} \cite{BPW}

The questions how
many orders of pertrubation theory are required in the calculation
of the effective hamiltonian for constituent quarks and gluons in QCD
and how large
must be the subspace of the light-front Fock space to diagonalize
the effective hamiltonian of QCD, require much more work to answer
than in the matrix model.
\ack{I am grateful to Pierre Grange for organizing the Les Houches
workshop on new light-front computational methods and to Robert Perry
for organizing the session on effective hamiltonians and renormalization
issues, and inviting me to speak. I am most indebted to Ken Wilson for 
discussions during my stay at The Ohio State University as a 
Fulbright Scholar in the academic year 1995/1996. It is my pleasure 
to thank Robert Perry for helpful comments and I would like 
to express my gratitude and thank him and Billy Jones, Martina Brisudov\'a 
and Brent Allen for discussions and hospitality extended to me at OSU. 
I have also discussed the subjects of my talk with Tomek Mas{\l}owski
and Marek Wi{\c e}ckowski.
Research described in this paper has been supported in part
by Maria Sk{\l}odowska-Curie Foundation under Grant No.  MEN/NSF-94-190.}


\begin{thebibliography}

\bibitem{GW1} 1. St. D. G{\l}azek, K. G. Wilson,
\review Phys. Rev. D, 48, 1993, 5863.

\bibitem{GW2} 2. St. D. G{\l}azek, K. G. Wilson,
\review Phys. Rev. D, 49, 1994, 4214.

\bibitem{W3} 3. K. G. Wilson et al,
\review Phys. Rev. D, 49, 1994, 6720.

\bibitem{WEG} 4. F. Wegner,
\review Ann. Physik, 3, 1994, 77.

\bibitem{WG} {5. K. G. Wilson, St. D. G{\l}azek,
in ``Computational Physics: Proceedings of the Ninth Physics Physics
Summer School at the Australian National University''; Eds. H. J. Gardner and
C. M. Savage, World Scientific, Singapore, 1997.}

\bibitem{GW3}{ 6. St. D. G{\l}azek, K. G. Wilson, ``Asymptotic Freedom and
Bound States in Hamiltonian Dynamics'', in preparation.}

\bibitem{G} {7. St. D. G{\l}azek, ``Renormalization of Hamiltonians
in the Light-Front Fock Space'', Warsaw University Report No. IFT/2/1997.}

\bibitem{J} 8. E.g. see R. Jackiw,
\book in ``M. A. B. B\'eg Memorial Volume'';
edited by A. Ali and P. Hoodbhoy, World Scientific, Singapore, 1991, 25.

\bibitem{Perry} {9. R. J. Perry,
``A Simple Confinement Mechanism for
Light-Front Quantum Chromodynamics'';
in ``Theory of Hadrons and Light-Front QCD'', Ed. St.D. G{\l}azek,
World Scientific, Singapore, 1995, p. 56, and references therein.
In particular, see the works by Perry and collaborators.}

\bibitem{BP} 10. M. Brisudov\'a, R. J. Perry,
\review Phys. Rev. D, 54, 1996, 1831.

\bibitem{BPW} 11. M.  Brisudov\'a, R. J.  Perry, K. G. Wilson,
\review Phys. Rev. Lett., 78, 1997, 1227.

\end{thebibliography}
\end{document}